# Critical Analysis of 5G Networks Traffic Intrusion using PCA, t-SNE and UMAP Visualization and Classifying Attacks

**Humera Ghani \*, Shahram Salekzamankhani, Bal Virdee**

Centre for Communications Technology, School of Computing and Digital Media, London Metropolitan University, London, N7 8DB, UK

hug0051@my.londonmet.ac.uk, s.salekzamankhani@londonmet.ac.uk, b.virdee@londonmet.ac.uk

**Abstract**   Networks, threat models, and malicious actors are advancing quickly. With the increased deployment of the 5G networks, the security issues of the attached 5G physical devices have also increased. Therefore, artificial intelligence based autonomous end-to-end security design is needed that can deal with incoming threats by detecting network traffic anomalies. To address this requirement, in this research, we used a recently published 5G traffic dataset, 5G-NIDD, to detect network traffic anomalies using machine and deep learning approaches. First, we analyzed the dataset using three visualization techniques: t-Distributed Stochastic Neighbor Embedding (t-SNE), Uniform Manifold Approximation and Projection (UMAP), and Principal Component Analysis (PCA). Second, we reduced the data dimensionality using mutual information and PCA techniques. Third, we solve the class imbalance issue by inserting synthetic records of minority classes. Last, we performed classification using six different classifiers and presented the evaluation metrics. We received the best results when K-Nearest Neighbors classifier was used: accuracy (97.2%), detection rate (96.7%), and false positive rate (2.2%).

## 1 Introduction

With the increased deployment of the 5G networks, the security issues of the attached 5G physical devices have also increased [1]. Several technologies, for example, firewalls, traffic shaping devices, and intrusion detection systems are available to secure a network [2]. With the changing world needs and threat models, networks are becoming more complex and heterogeneous; malicious actors are also becoming more advanced [3]. Hence, artificial intelligence based autonomous end-to-end security design is needed that can deal with incoming threats by detecting network traffic anomalies [4]. Hence, to address network traffic anomaly issues in 5G networks, we proposed a novel approach using the recently released 5G traffic dataset, 5G-NIDD. We used machine and deep learning approaches to perform our experiments.

In this study we first performed a visual analysis of this dataset using three different visualization techniques. We then reduced the data dimensionality. In this step, feature selection and feature extraction were performed using mutual information and principal component analysis techniques. In the third step, the class imbalance issue was resolved by inserting synthetic records of minority classes. We used a random over-sampling method for balancing the class distribution. Finally, we performed classification using six different classifiers and presented the evaluation metrics. The best results were for K-Nearest Neighbors classifier with an accuracy of 97.2%, detection rate of 96.7%, and false positive rate of 2.2%.

The contributions of this paper are:
- Visual analyses of the 5G-NIDD dataset to understand its intricacies better by using PCA, t-SNE, and UMAP techniques.

- Dimensionality reduction to remove unimportant features, which cause inaccurate results, more processing time and high computational power, using information gain and principal component analysis techniques.
- Remove class imbalance to improve classification metrics using the random over sampling technique.
- Classify benign and malicious traffic with high accuracy using decision tree (DT), k-nearest neighbors (KNN), multi-layer perceptron (MLP), naïve bayes (GNB), random forest (RF), and support vector classifier (SVC) algorithms.

This paper is structured as follows. Section 2 describes the related work reported recently in the literature. Section 3 introduces the dataset. Section 4 elaborates on the proposed approach. Section 5 shows the results and discusses findings. Section 6 concludes the work presented in this paper and recommends future work.

## 2 Related Works

This section discusses contemporary research in the field of network anomaly detection. Researchers are addressing this problem using various machine and deep learning approaches. For experiments, in general, they employ CICIDS-2017, NSL-KDD, and UNSW-NB15 datasets. However, we used the 5G-NIDD dataset, a comparatively new dataset having 5G network traffic records.

Authors [5] divided the UNSW-NB15 dataset based on protocol: TCP, UDP, and OTHER. Using the Chi-square technique, they performed feature selection, and for classification, they used a 1-dimensional convolution neural network. Their work includes the visualization of the dataset using t-SNE. However, current research [7, 8] have reported better classification performance metrics on the same dataset.

Authors in [6] created and evaluated the 5G-NIDD dataset using five different classifiers on binary and multiclass labels. They used the analysis of variance (ANOVA) technique for feature selection and used the ten best features for classification. Features in this dataset have skewness and multi-modal properties, while ANOVA is used for features having a normal distribution. Therefore, using ANOVA on this dataset for feature selection is inappropriate.

Reference [7] performed their experiments on NSL-KDD and UNSW-NB15 datasets. They proposed combining particle swarm optimization (PSO) and a gravitational search algorithm (GSA) for feature selection. They used five other feature selection techniques but received the best classification metrics when the features selected by their proposed method were given to the random forest classifier. Although this research received good results, their proposed technique selected a high number of features in comparison to the other feature selection methods [5] evaluated in this paper.

Researchers in [8] did their experiments on UNSW-NB15, CICIDS-2017, and Phishing datasets. They used a correlation-based feature selection method. For data visualization and feature reduction, they used t-SNE and for the classification, random forest. But, [9] mentioned a limitation that t-SNE is a non-parametric dimensionality reduction technique; these techniques cannot map new data points.

Reference [10] used NSL-KDD dataset for their research. They used an auto-encoder for detecting network anomalies. Although they reported good classification performance, contemporary research [7] showed better performance metrics on the same dataset.

Authors in [11] used UNSW-NB15 and NSL-KDD datasets to investigate network traffic anomalies. First, they address the class imbalance issue by reducing the noise samples from the majority class then they increase the minority class samples using Synthetic Minority Over-sampling Technique. Second, they performed classification using deep learning approaches: Convolution Neural Network and Bi-directional long short-term memory.

Researchers in [12] performed their experiments on UNSW-NB15 and NSL-KDD (KDDTest+ and KDDTest-21) datasets. First, they addressed the class imbalance issue using Wasserstein Generative Adversarial Network. Second, they employed a Stacked Autoencoder for feature extraction. Third, they constructed a cost-sensitive loss function. Their performance metrics suggest further improvement in their approach.

The above discussion clarifies that current research in network traffic anomaly detection lacks in presenting the visual analysis of datasets. Therefore, this research visually examined 5G-NIDD dataset, a newly released 5G traffic dataset [6].

## 3 Dataset

5G-NIDD dataset is created using a real 5G test network for network intrusion detection. It was published by [6]. It has total 52 features, 32 float type, 12 int type and 8 categorical type. This dataset has 1215890 records, where 477737 are benign, and 738153 are malicious. Benign records are 39.2%, whereas malicious records are 60.7% as shown in Table 1. This dataset has eight different types of attacks; their names and percentage in the attack traffic are: UDPFlood (61.9%), HTTPFlood (19.0%), SlowrateDoS (9.9%), TCPConnectScan (2.7%), SYNScan (2.7%), UDPScan (2.1%), SYNFlood (1.3%), and ICMPFlood (0.15%) Table 2.

**Table 1.** Distribution of records in 5G-NIDD dataset

| Label | No. of records | Percentage |
|---|---|---|
| Benign | 477737 | 39.291 |
| Malicious | 738153 | 60.708 |
| Total | 1215890 | 100 |

**Table 2.** Distribution of attacks in malicious records

| Attack Type | No. of records | Percentage |
|---|---|---|
| UDPFlood | 457340 | 61.957 |
| HTTPFlood | 140812 | 19.076 |
| SlowrateDos | 73124 | 9.9063 |
| TCPConnectScan | 20052 | 2.716 |
| SYNScan | 20043 | 2.715 |
| UDPScan | 15906 | 2.154 |
| SYNFlood | 9721 | 1.316 |
| ICMPFlood | 1155 | 0.156 |
| Total | 738153 | 100 |

## 4 Proposed Approach

This section describes the approach adopted to perform this research. First, data cleaning and wrangling were performed at the data preprocessing stage. Second, a visual analysis of data was presented and described using three different methods. Third, data dimensionality was reduced by employing feature selection and feature extraction approaches. Fourth, the class imbalance issue was addressed. Fifth, traffic classification was performed using six different classification algorithms. Finally, evaluation metrics were described and results were presented.

## 4.1 Data Pre-processing

Data is prepared to feed into the machine learning algorithm. This dataset had redundant and unnecessary features; therefore, data denoising was performed to remove redundant and unnecessary features. Some features had null values, which were imputed with appropriate alternative values. Some features showed skewness and multimodal distribution; therefore, log transformation was performed to achieve normal distribution. Categorical features were encoded using One-hot encoding method. The dataset was split into train and test sets to avoid overfitting. Lastly, it was normalized to have mean 0 and variance 1.

## 4.2 Visual Analysis of 5G-NIDD dataset

In this paper, visualization is performed to understand the data better so researchers can apply appropriate processing techniques to achieve good results. This research used three visualization techniques to get deep insight into the dataset. These techniques are PCA, t-SNE, and UMAP. PCA is one of the widely used dimensionality reduction techniques that performs orthogonal linear transformation of correlated variables into uncorrelated features. These new features are called principal components. This technique works to preserve the variance of original high-dimensional data into low-dimensional principal components. t-SNE is a non-linear dimensionality reduction and visualization technique that creates probability distribution of closest points. This algorithm tries to transform high-dimensional feature space into 2-dimensional feature space by minimizing the difference between two distributions. It models high-dimensional space into Gaussian distribution while 2-dimensional space into t-distribution. UMAP is an efficient dimensionality reduction technique that uses a graph algorithm to reduce data dimensionality. First, it constructs the topology of the high-dimensional data. Then it constructs low-dimensional local clustering by grouping similar observations.

The class distribution of 5G-NIDD dataset is imbalanced (see Figure 1). Three attack categories (UDPFlood, HTTPFlood, and SlowrateDos) constitute 90.8% malicious traffic, while the other five attack categories (TCPConnectScan, SYNScan, UDPScan, SYNFlood, and ICMPFlood) constitute 8.95% malicious traffic Table 2. Figure 2 and Figure 3 show a 2-dimensional and 3-dimensional view of principal components; the class imbalance issue can be observed clearly.

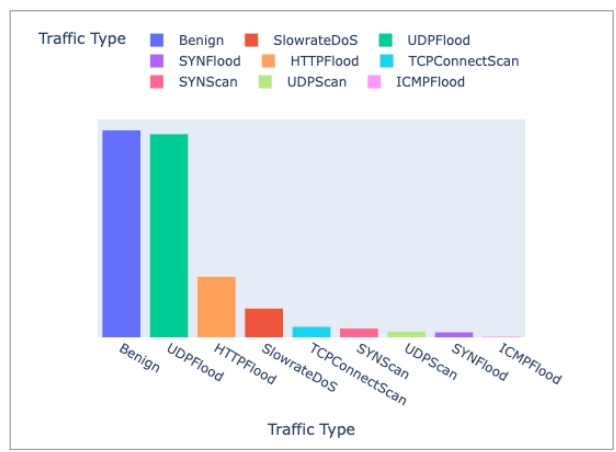

**Fig. 1.** Distribution of traffic types in the 5G-NIDD dataset

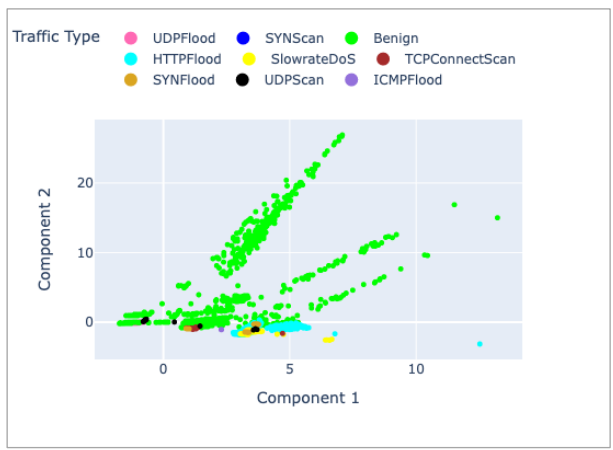

**Fig. 2.** 2D plot of PCA showing all type of traffic

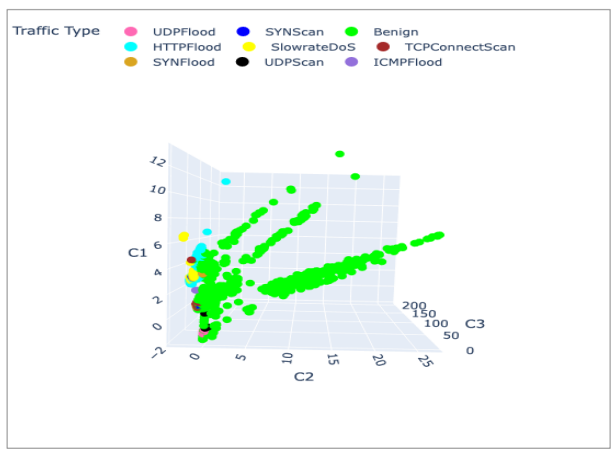

**Fig. 3.** 3D plot of PCA showing all type of traffic

Figures 4 and 5 show the complete datasets using t-SNE and UMAP, respectively. In addition to class imbalance, class overlap and within-class clustering issues are evident from these figures.

The class overlap issue is confirmed in Figures 6-9, where HTTPFlood and SlowrateDoS classes are overlapping. All three visualization techniques confirm this issue. Figures 6 and 7 show 2-dimensional and 3-dimensional plots of class overlap using principal components. Figure 8 shows a t-SNE plot where the class overlap is evident. Figure 9 shows a UMAP plot showing class overlap.

This dataset has within-class cluster issues as well. Figures 6-9 show scattered clusters of HTTPFlood and SlowrateDoS classes using PCA, t-SNE, and UMAP visualizations.

Class imbalance and overlap can hinder the attack detection performance [13, 14]. In this research, we addressed the class imbalance problem. Our work achieved good evaluation metrics (see Table 4), suggesting that if a dataset has a high number of records, classifiers can identify the classes with accuracy despite the class overlap issue.

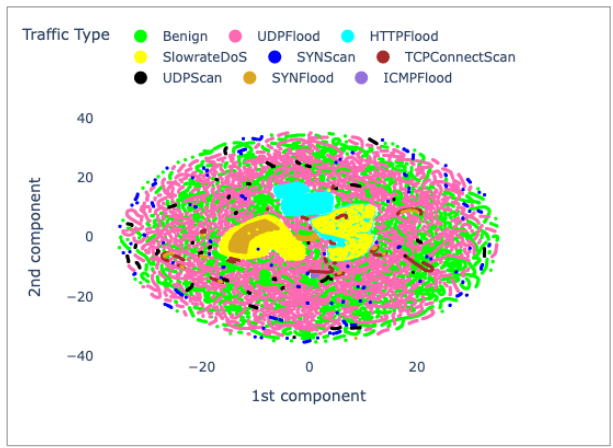

**Fig. 4.** Full dataset visualization (t-SNE)

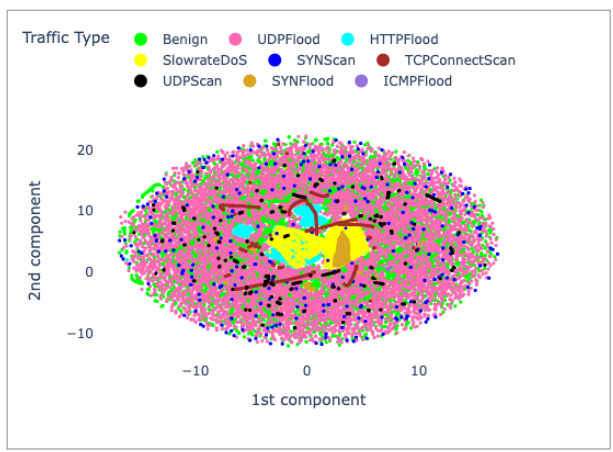

**Fig. 5.** Full dataset visualization (UMAP)

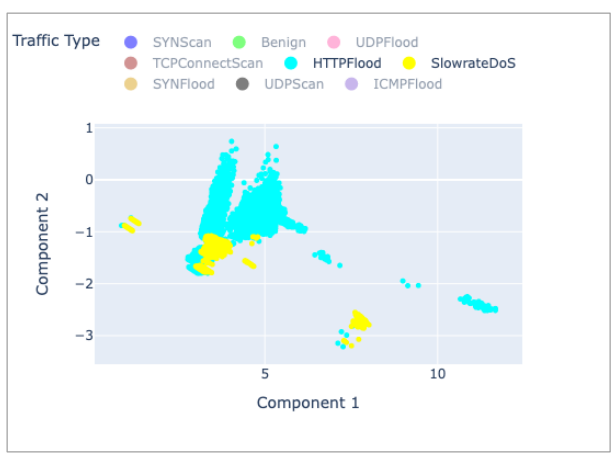

**Fig. 6.** HTTPFlood and SlowrateDoS overlap (2D PCA)

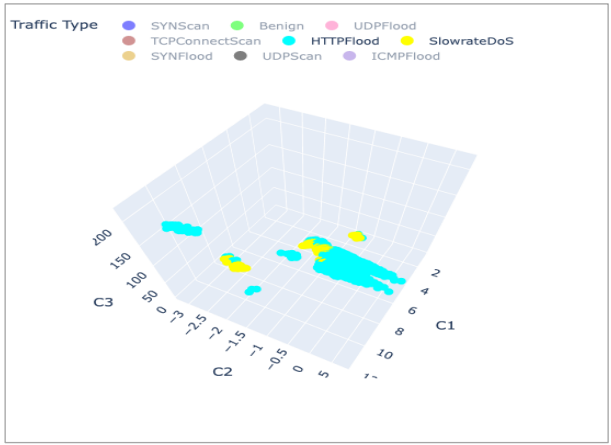

**Fig. 7.** HTTPFlood and SlowrateDoS overlap (3D PCA)

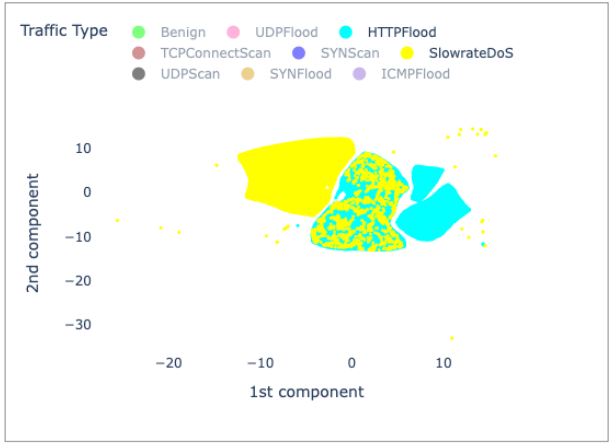

**Fig. 8.** HTTPFlood and SlowrateDoS overlap (t-SNE)

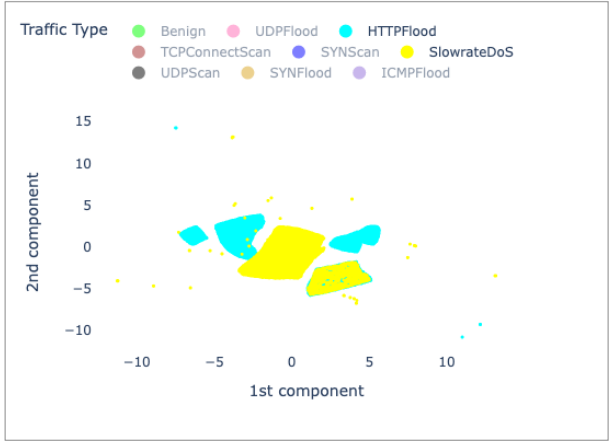

**Fig. 9.** HTTPFlood and SlowrateDoS overlap (UMAP)

### 4.3 Dimensionality Reduction

Dimensionality Reduction transforms high dimensional data into low dimensions where newly transformed data is a meaningful representation of original data [15]. Dimensionality reduction techniques are employed to reduce the number of variables, reduce the computational complexity of high-dimensional data, improve the model accuracy, better visualization, and understand the process that generated the data [16]. Two main approaches of dimensionality reduction are feature selection and feature extraction.

Feature Selection methods select the most valuable features from the feature space. This process creates a low-dimensional feature space representation that preserves the most valuable information. [16] identified two approaches to feature selection: the filter method and the wrapper method. In this paper, the most useful features are selected using one of the filter method algorithms, Mutual Information. This algorithm measures the amount of information that a random variable contains another random variable, in other words, the reduction of the uncertainty of the original random variable, given the knowledge of another random variable [17]. Unlike Pearson correlation, this method can measure the nonlinear relationship between two variables.

Feature extraction is producing a compressed representation from the input vector. Feature extraction techniques create new features from the original data space using functional mapping [18]. Several algorithms are available that can perform this transformation linearly and non-linearly. Researchers in [16] identified three approaches to feature extraction: performance measure, transformation, and generation of new features. We choose the transformation technique PCA as it is faster since the first few principal components are computed and more interpretable than other techniques, such as Auto-encoder.

### 4.4 Remove Class Imbalance

Class imbalance problem occurs when some classes have more instances than others; in such cases, learning algorithms are overwhelmed by the large classes and ignore the small classes [19]. Learning algorithms are generally not designed to handle imbalanced datasets without proper adjustment [20]. Researchers in [21] pointed datasets frequently exhibit class imbalance and overlap issues. 5G-NIDD dataset also shows class imbalance in Figure 1.

There are several approaches to solving the class imbalance problem. These approaches can be grouped as: data-level, cost-sensitive, and ensemble learning. Data-level approaches modify the dataset. Cost-sensitive approaches modify the cost that algorithm tries to optimize. The ensemble learning approach leverages the power of several learners to predict the minority class. Data-level approach Random Over-sampling increases the number of observations from the minority class at random. In contrast, Random Under-sampling decreases the number of observations from the majority class at random.

Synthetic Minority Over Sampling (SMOTE) technique is frequently employed in contemporary research [11, 22, 23] to overcome class imbalance issues from network intrusion detection datasets. It balances class distribution by randomly inserting minority class samples. It does linear interpolation to produce synthetic records of the minority class. These records are created by selecting k-nearest neighbors for each example in the minority class. We chose SMOTE to solve the class imbalance issue in the dataset.

### 4.5 Network Traffic Classification

We used tree-based, probability-based, proximity-based, deep learning, and support vector classifiers to predict class labels.

Decision Tree is a non-parametric learning algorithm. It works on a divide-and-conquer strategy. This greedy algorithm searches and identifies optimal split points within a tree. It does this job reclusively, which completes when most records classify under the same class label.

k-Nearest Neighbor is a proximity-based classifier. To predict the class label of a point, first, it finds K nearest neighbors of this point based on Euclidean distance. Then, each of these neighbors votes for their class, and the majority class wins.

Multilayer Perceptron is a powerful deep-learning model which is inspired by neurons in the human brain. The basic building blocks of MLP are perceptrons which are simple processing units. It can have many layers of perceptrons, which gives it the name MLP.

Naïve Bayes classier is simple and fast, has very few tunable parameters, and good for high dimensional data. Given the value of class variable this algorithm assumes conditional independence between each pair of input variables.

Random Forest classifier is an ensemble of decision trees that can perform classification using a majority vote. Each decision tree uses a randomly selected feature set from the original dataset. In addition, each tree uses a different sample of data, like the bagging approach. It can successfully model high-dimensional data where features are non-linearly related, and it doesn't assume data follows a particular distribution.

Support Vector Classifier is a simple and powerful classifier. It can draw linear and non-linear class boundaries to classify the data points. To perform its job, it iteratively constructs a hyper-plane to differentiate classes. Each iteration tries to minimize the error. The main idea of this technique is to create a hyper-plane that can best divide the data into classes.

### *4.6 Evaluation Metrics*

Model performance is evaluated using accuracy, detection rate, and false positive rate metrics. Elements of these metrics can be retrieved from the confusion matrix where the confusion matrix is {TP, TN, FP, FN}. True Positive (TP) means correctly classified attack packets. True Negative (TN) means correctly classified normal packets. False Positive (FP) means incorrectly classified attack packets and False Negative (FN) means incorrectly classified normal packets.

Accuracy means the ratio between correctly identified packets and total number of packets.
$$\text{Accuracy} = \frac{(TP+TN)}{(TP+TN+FP+FN)} \tag{1}$$

Detection Rate represents the ratio of correctly identified attacks versus predicted attacks.
$$\text{Detection Rate} = \frac{TP}{(TP+FN)} \tag{2}$$

False Positive Rate is the ratio of incorrectly identified attacks versus predicted normal.
$$\text{False Positive Rate} = \frac{FP}{(FP+TN)} \tag{3}$$

## 5 Results and Discussion

To reduce the feature space, we used mutual information and PCA techniques. We processed the dataset before applying dimensionality reduction techniques (see Section 4.1). Mutual Information technique ranked the features (see Figure 10). We selected twenty-two top-ranked features and transformed them into eleven principal components. These eleven components captured 89.2% variance in the data (see Table 3).

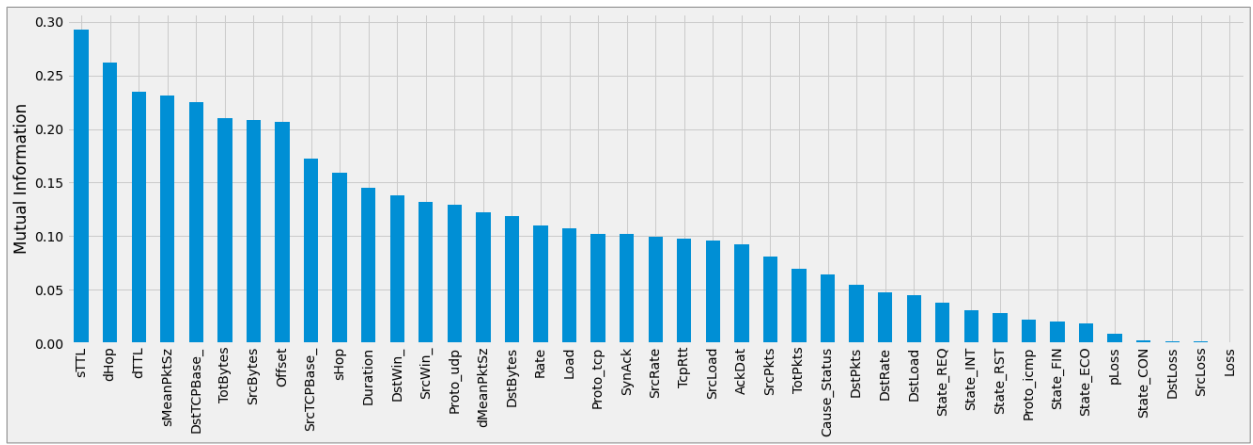

**Fig. 10.** Mutual information ranking of features

**Table 3.** PCA explained variance ration

| Principal Component | Variance Captured |
| --- | --- |
| 1st component | 0.20667303 |
| 2nd component | 0.14651562 |
| 3rd component | 0.09805928 |
| 4th component | 0.08302226 |
| 5th component | 0.05968254 |
| 6th component | 0.05411584 |
| 7th component | 0.05063291 |
| 8th component | 0.05041822 |
| 9th component | 0.04955088 |
| 10th component | 0.04887834 |
| 11th component | 0.04494785 |
| Total | 0.89249676 |

Eleven principal components are fed to classifiers for classification. Table 4 shows the classification performance of six classifiers using accuracy, detection rate, and false positive rate. Our results show that GNB classifier showed considerably low-performance metrics. This result confirms [6] finding. DT, RF, and kNN showed better performance metrics than MLP and SVC. However, kNN remains the best classifier in all evaluation metrics with accuracy (97.2%), detection rate (96.7%), and false positive rate (2.2%). Figure 11 shows ROC curve of k-NN classifier that captured 97.2% area under the curve.

Table 5 shows a comparison of our approach with other contemporary techniques. Research [11] reported 77.16% and 83.58% accuracy on UNSW-NB15 and NSL-KDD datasets, respectively. Research [12] showed 93.27% and 90.24% accuracy on UNSW-NB15 and NSL-KDD datasets, respectively. Our approach outperformed them and achieved 97.28% classification accuracy using the 5G-NIDD dataset.

**Table 4.** Evaluation Metrics (binary classification)

| Evaluation Metrics | DT | RF | KNN | GNB | MLP | SVC |
|---|---|---|---|---|---|---|
| Accuracy | 97.150 | 97.168 | 97.275 | 87.425 | 94.91 | 92.09 |
| Detection Rate | 96.618 | 96.650 | 96.765 | 92.428 | 92.646 | 90.49 |
| False Positive Rate | 2.318 | 2.314 | 2.213 | 17.559 | 2.828 | 6.32 |

**Table 5.** Performance comparison

| Research | Accuracy | Dataset |
|---|---|---|
| [11] | 77.16% | UNSW-NB15 |
|  | 83.58% | NSL-KDD |
| [12] | 93.27% | UNSW-NB15 |
|  | 90.34% | NSL-KDD |
| Proposed Research | 97.28% | 5G-NIDD |

A limitation of this research is that the 5G-NIDD dataset was published recently [6]. Yet, more research needs to be produced on it to compare our results. Another limitation is the processing power. If it were not the case, we would have searched for the best hyperparameters for our classifiers and reported even better evaluation metrics.

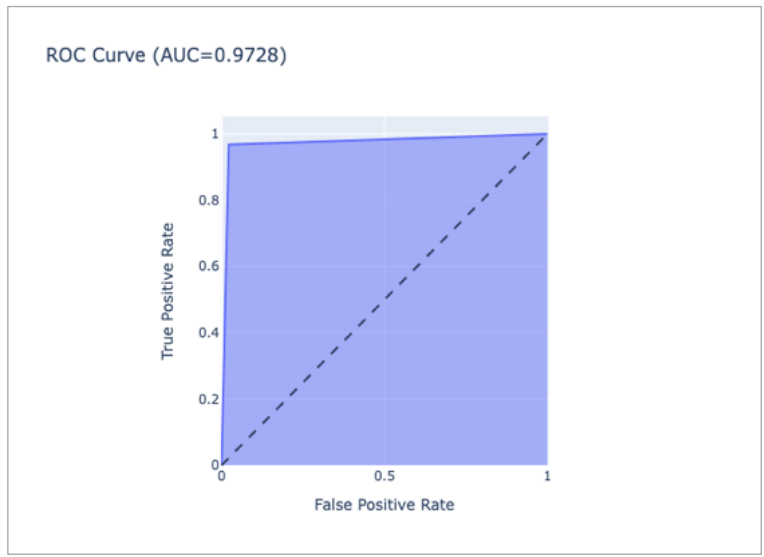

**Fig. 11.** ROC curve of k-NN

## 6 Conclusion and Future Work

In this paper, we presented a novel approach to classify network traffic anomalies with high accuracy. We analyzed the dataset by projecting it in 2-dimensional and 3-dimensional spaces using linear and non-linear dimensionality reduction and visualization techniques. We reduced the feature space by ranking them using the Mutual Information algorithm then we transformed high-ranked features into principal components. This dataset had a class imbalance issue; we solved it by balancing the class distribution using a random oversampling algorithm. Last, we performed classification using six classification algorithms and presented the evaluation metrics using accuracy, detection rate, and false positive rate. We achieved the best classification performance when the k-nearest neighbors algorithm was used.

In the future, we intend to extend this research in two directions: use an ensemble learner to improve classification metrics and use a generative model for class imbalance issues since it is one of the highly successful deep learning architectures.